\documentstyle[preprint,aps,prl]{revtex}

\begin{document}

\draft

\title{Critical temperature and condensate fraction 
of the trapped interacting Bose gas  with finite-size effects} 

\author{ Hongwei Xiong$^{1,2}$, Shujuan Liu$^{1}$, Guoxiang Huang$^{3}$, Zhijun Xu$^{1}$, Cunyuan Zhang$^{1}$}

\address{$^{1}$Department of Applied Physics, Zhejiang 
University of Technology, Hangzhou, 310032, P. R. China} 
\address{$^{2}$Zhijiang College,  
Zhejiang University of Technology, Hangzhou, 310012, P. R. China }
\address{$^{3}$Department of Physics, East China Normal University, 200062, Shanghai, P. R. China}

%\date{}

\maketitle

\begin{abstract}

{\it The critical temperature and condensate fraction of a trapped
interacting Bose gas are investigated when both atom-atom interaction and
finite-size effects are taken into account. Canonical ensemble is used to obtain
the equations
on the condensate fraction for the trapped interacting Bose gas
near and below the critical temperature.
In our approaches corrections due to atom-atom
interaction and finite-size effects
are obtained simultaneously for the critical temperature and
condensate fraction of the system. Analytical high-order correction to the
condensate fraction is given in this work.
} 

\end{abstract}

\pacs{ 03.75.Fi, 67.40.Kh, 05.30.Jp}

\narrowtext

%%%%%%%%%%%%%%%%%%%%%%%%%
\section{Introduction}
%%%%%%%%%%%%%%%%%%%%%%%%%

The experimental realization of Bose-Einstein condensation (BEC) 
in the dilute alkali-metal atoms \cite{EXP} and more recently in the 
atomic hydrogen \cite{MIT} has stimulated a new interest in the 
theoretical study of the inhomogeneous Bose gas. 
Thermodynamic properties such as critical
temperature, condensate fraction have been investigated by 
several authors
for the trapped Bose gas recently \cite{RMP}. 

In the presence of an anisotropic harmonic potential of the form
${V_{ext}(%
\vec{r})=m(\omega _{x}^{2}x^{2}+\omega _{y}^{2}y^{2}+\omega _{z}^{2}z^{2})/2}$
the noninteracting model gives the critical temperature
${T_{c}^{0}=\frac{%
\hbar \omega _{ho}}{k_{B}}\left( \frac{N}{\zeta (3)}\right) ^{1/3}}$ \cite{RMP}, 
where  
$\omega _{ho}=\left( \omega _{x}\omega _{y}\omega _{z}\right) ^{1/3}$ 
is the geometrical average of the oscillator frequencies.
$\zeta (n)$ is Riemann $\zeta$ function. 
In the large-N limit, the
condensate fraction is $\xi =\frac{N_{0}}{N}=$ ${1-\frac{\zeta (3)}{N}\left( 
\frac{k_{B}T}{\hbar \omega _{ho}}\right) ^{3}}$.
Finite-size effects \cite{FIN}
and interatomic interaction will give corrections to the thermodynamic
properties of the system. The correction to the critical temperature 
due to finite-size effects has shown to be 
$\delta {T_{c}^{0}/T_{c}^{0}\simeq -0.73}\overline{\omega }N^{-1/3}/\omega_{ho}$,
where $\overline{\omega }%
=(\omega _{x}+\omega _{y}+\omega _{z})/3$ is the mean frequency.

Although the atom clouds realized in the present experiments are very
dilute, the effects due to interatomic interaction are important at low temperature.
Researches show that atom-atom interaction will give
 leading corrections when $N$ is larger than $10^{5}$ for the alkalis.
In fact the
question of how two-body forces affect the thermodynamic properties
of these
systems have been the object of several theoretical investigations
\cite{BEF}. Using a local density approximation 
Giorgini, Pitaevskii, and
Stringari \cite{RMP,GIO}
obtained the shift of the critical 
temperature
due to interatomic interaction: 
$\delta {T_{c}^{0}/T_{c}^{0}\simeq -1.33}%
aN^{1/6}/a_{ho}$, where $a_{ho}=\sqrt{\hbar /m\omega _{ho}}$ 
is the harmonic oscillator length. Recently Monte Carlo simulation \cite{MCS} 
is also used to investigate the thermodynamic properties of the 
trapped interacting Bose gases.

In this work, we investigate the critical temperature and the 
condensate fraction of the system when both atom-atom interaction and 
finite-size effects are taken into account. Canonical ensemble 
is used to obtain the analytical high-order correction due to interatomic interaction. 
Especially, finite-size effects are obtained simultaneously.
To obtain the shift of the critical temperature, we give the analytic description of
the condensation fraction near the critical temperature. 

The paper is planed as follows. In Sec.II we outline the 
canonical ensemble.
In Sec.III we give the condensate 
fraction of the system near the critical temperature. The shift of the critical 
temperature agrees with the usual results \cite{RMP,FIN,GIO}.
In Sec.IV we obtain the analytical high-order correction 
to the condensate fraction due to atom-atom interaction. 
The lowest-order correction agrees with the well-established results \cite{RMP,GIO,NAR}.
In Sec.V the correction to the condensate fraction beyond mean field theory is given.

%%%%%%%%%%%%%%%%%%%%%%%%%%%%%%%%%%%%%%%
\section{Partition Function of the Trapped Interacting Bose Gases}
%%%%%%%%%%%%%%%%%%%%%%%%%%%%%%%%%%%%%%%

Canonical ensemble and saddle-point method have been used to investigate the 
thermodynamic properties of the interacting uniform Bose gases \cite{PAT}.
In this work canonical 
ensemble is used to discuss the trapped interacting Bose gases.
The partition function of $N$ trapped interacting bosons is given by

\begin{equation}
{Q(N)=\Sigma _{n}\exp \left( -\beta E_{n}\right) =\Sigma _{\left\{
n_{x},n_{y},n_{z}\right\} }\exp \left[ -\beta \left\{ \Sigma
_{n_{x},n_{y},n_{z}}N_{n_{x}n_{y}n_{z}}\varepsilon
_{n_{x}n_{y}n_{z}}+E_{int}\right\} \right] }, 
\label{par1}
\end{equation}

\noindent where $N_{n_{x}n_{y}n_{z}}$ and
$\varepsilon _{n_{x}n_{y}n_{z}}$ are the
occupation numbers and energy level of the state 
$\{n_{x},n_{y},n_{z}\}$
respectively. $E_{int}$ is the interaction energy of the system.

From (\ref{par1}) 

\begin{equation}
Q(N){=\Sigma _{N_{0}=0}^{N}}\left\{ {\exp }\left[ -\beta \left(
E_{0}+E_{int}\right) \right] {Q}_{0}\left( N-N_{0}\right) \right\}, 
\label{par2}
\end{equation}

\noindent where ${Q}_{0}\left( N-N_{0}\right) ={\Sigma _{\{n_{x},n_{y},n_{z}\}}^{\prime
}\exp \left[ -\beta \left( \Sigma
_{n_{x},n_{y},n_{z}}N_{n_{x}n_{y}n_{z}}\varepsilon _{n_{x}n_{y}n_{z}}\right)
\right] }$ 
stands for the partition function of a fictitious system of 
$N-N_{0}$ trapped noninteracting atoms. 
$E_{0}$ is the energy of the condensate.
For convenience, 
we have separated
out the ground state $n_{x}=n_{y}=n_{z}=0$ from the states 
$\left\{n_{x},n_{y},n_{z}\right\} \neq 0$, 
so that we first carry out the primed
summation over all 
$\left\{ n_{x},n_{y},n_{z}\right\} $ ($\left\{
n_{x},n_{y},n_{z}\right\} \neq 0$) 
with a fixed value $(N-N_{0})$ of the
partial sum $\Sigma _{\left\{ n_{x}n_{y}n_{z}\right\} }^{\prime }$ 
and then
carry out a summation over all possible values of $N_{0}$, 
{\it ie.} $N_{0}=0$ to $N_{0}=N$. 
Assume $A_{0}\left( N-N_{0}\right) $ is the free energy of the
fictitious system. 
$A_{0}(N-N_{0})=-\frac{1}{\beta }\ln Q_{0}(N-N_{0})$. From (\ref{par2})

\begin{equation}
Q(N){=\Sigma _{N_{0}=0}^{N}\exp }\left[ -\beta \left( E_{0}+E_{int}\right)
-\beta A_{0}(N-N_{0})\right].
\label{par3}
\end{equation}

The sum ${\Sigma _{N_{0}=0}^{N}}$ in (\ref{par3})
maybe replaced by the largest
term in the sum, for the error omitted in doing so will be 
statistically negligible. With this approximation we do not investigate
the fluctuations of the system.
However, this approximation is reasonable
because we can give the occupation number of the energy level 
$\varepsilon _{n_{x}n_{y}n_{z}}$,
which agrees with the widely used 
mean occupation number in the frame of grand-canonical ensemble.
Neglecting all terms but the largest in (\ref{par3}), the number of bosons in
the condensate can be obtained as:

\begin{equation}
-\beta \frac{\partial }{\partial N_{0}}\left( E_{0}+E_{int}\right) -\beta 
\frac{\partial }{\partial N_{0}}A_{0}(N-N_{0})=0.
\label{eq4}
\end{equation}

The calculations of the free energy $A_{0}(N-N_{0})$ of 
the fictitious noninteracting Bose gas is nontrivial because there 
is a requirement that the number of the particles is $N-N_{0}$ 
in the summation of the partition function. Using the 
saddle-point method of integration
developed by Darwin and Fowler \cite{DAR} it is straightforward to find that
$-\beta \frac{\partial }{\partial N_{0}}A_{0}(N-N_{0})=\ln z_{0}$
in which $z_{0}$ is the fugacity of the fictitious Bose gas. In addition, the fugacity $z_{0}$ is
determined by

\begin{equation}
{N-N}_{0}{=\Sigma _{n_{x},n_{y},n_{z}}^{\prime}\frac{1}{\exp \left[ \beta
\varepsilon _{n_{x}n_{y}n_{z}}\right] z_{0}^{-1}-1}}. 
\label{main1}
\end{equation}

We can easily understand (\ref{main1})
in terms of Bose-Einstein distribution of the
trapped noninteracting Bose gas. Using the relation 
$-\beta \frac{\partial }{\partial N_{0}}A_{0}(N-N_{0})=\ln z_{0}$, (\ref{eq4}) becomes

\begin{equation}
-\beta \frac{\partial }{\partial N_{0}}\left( E_{0}+E_{int}\right) 
+lnz_{0}=0. 
\label{main2}
\end{equation}

Equations (\ref{main1}) and (\ref{main2}) will be used to discuss the corrections due to
atom-atom interaction and finite-size effects.
Once we know the interaction energy of the system, it is
easy to obtain the correction due to atom-atom interaction.
In addition, finite-size effects are separated off in (\ref{main1}).
Omitting interactions between atoms, we can obtain $\ln z=\beta \epsilon _{000}$ 
from (\ref{main2}). From (\ref{main1}) the
number of the thermal atoms $N_{T}$ is given by

\begin{equation}
N_{T}={N-N}_{0}{=\Sigma _{n_{x},n_{y},n_{z}}^{\prime}}\frac{1}{\exp \left[\beta
(\varepsilon _{n_{x}n_{y}n_{z}}-\epsilon _{000})\right] -1}. 
\label{eq7}
\end{equation}

This is the exact conclusion in the frame of grand-canonical ensemble for the
trapped noninteracting Bose gas. This shows the equivalence between the 
grand-canonical ensemble and canonical ensemble in the calculations of the
condensate fraction. In addition, it shows that neglecting all terms but the largest
in (\ref{par3}) is reasonable.
In the absence of interaction, from (\ref{eq7}),
the condensate fraction is given by \cite{FIN}

\begin{equation}
\xi =1-(\frac{T}{T_{c}^{0}})^{3}-\frac{3\overline{\omega }\zeta \left( 2\right) }{2\omega _{ho}\zeta \left( 3\right) ^{2/3}}(\frac{T}{T_{c}^{0}})^{2}N^{-1/3}.
\label{eq8}
\end{equation}

The third term in (\ref{eq8}) is the finite-size correction to the ideal Bose gas in the 
thermodynamic limit. In short, (\ref{main1}) and (\ref{main2}) account for both interatomic
interaction and finite-size effects. We will use (\ref{main1}) and (\ref{main2}) 
to calculate the critical temperature and condensate fraction of the system.

%%%%%%%%%%%%%%%%%%%%%%%%%%%%%%%%%%
\section{critical temperature and condensate fraction near $T_{c}$}
%%%%%%%%%%%%%%%%%%%%%%%%%%%%%%%%%

%%%%%%%%%%%%%%%%%%%%%%%%%%%%%%%%%%%
\subsection{interaction energy of the system near the critical temperature}
%%%%%%%%%%%%%%%%%%%%%%%%%%%%%%%%%%%

The parameter expressing the importance of the interatomic interaction 
compared to the kinetic energy is 
$\frac{E_{int}}{E_{kin}} \propto \frac{N_{0}|a|}{a_{ho}}$, 
where $a$ is the scattering length between bosons. Near the critical 
temperature $\frac{N_{0}|a|}{a_{ho}}<<1$.  This means that $E_{kin}>>E_{int}$.
In this case we can use the method of
pseudopotentials developed by Huang and Yang \cite{YAN}
to calculate the interaction energy of the system.

In the method of pseudopotentials,
the actual Hamiltonian of the system is replaced by an
effective Hamiltonian, such that the ground state and the low-lying
energy
levels of the system are given equally well by 
the new Hamiltonian. With
hard-sphere approximation, the boundary conditions 
between atoms are
replaced by the pseudopotential operator 
$g\delta
^{(3)}(\vec{r})\frac{\partial }{\partial r}r$ 
in the effective Hamiltonian, where $g=\frac{4\pi a\hbar^{2}}{m}$.
The effective Hamiltonian of the system may be taken to be

\begin{equation}
\widehat{H}=-\frac{\hbar ^{2}}{2m}\left( \nabla _{1}^{2}+\cdots +\nabla
_{N}^{2}\right) +\Sigma _{i=1}^{N}V_{ext}(\vec{r}_{i})+\Sigma _{i<j}\omega
_{ij},
\label{eff}
\end{equation}

\noindent where $\omega _{ij}=g\Sigma _{i<j}\delta
^{(3)}\left( \vec{r}_{i}-\vec{r}_{j}\right) \frac{\partial }{\partial r_{ij}}%
r_{ij}$. For a trapped dilute Bose gas near the critical temperature
the pseudopotentials can be regarded as perturbation terms. The energy
levels to the first order in $a$ may be obtained through the usual perturbation theory.

In the general case of a system containing an arbitrary number $N$ of
bosons, the unperturbed normalized wave function of the system is given by

\begin{equation}
\Psi _{N}=\left( \frac{N_{\alpha _{1}}!N_{\alpha _{2}}!\cdots }{N!}\right)
^{1/2}\Sigma_{P}\lbrack \phi _{\alpha _{1}}\phi _{\alpha
_{2}}\cdots \phi _{\alpha _{N}}\rbrack,
\end{equation}

\noindent where the sum is taken over all permutations of the different suffixes $%
\alpha _{1}$, $\alpha _{2}$, $\cdots $, $\alpha _{N}$ and the numbers $%
N_{\alpha _{i}}$ show how many of these suffixes have the same value $\alpha
_{i}$ (with $\Sigma N_{\alpha _{i}}=N$). 
From the first-order perturbation theory the interaction energy $E_{int}$ of the system is 
given by

\begin{equation}
E_{int}=<\Psi _{N}|\Sigma _{i<j}\omega _{ij}|\Psi _{N}>.
\end{equation}

Thus, the operators $\left( \partial /\partial r_{ij}\right) r_{ij}$ will be
operating on a set of functions which are well behaved for all values of $%
r_{ij}$; accordingly, these operators may be replaced by the unit operators.
Basing on the investigation of the various permutations \cite{PAT} of the single
particle states the interaction energy takes the form, 

\begin{equation}
E_{int}=\frac{g( N_{0}^{2}-N_{0})}{2}\int \phi
_{0}^{4}(\vec{r})d^{3}\vec{r}+2gN_{0}\int \phi
_{0}^{2}(\vec{r})n_{T}\left( \vec{r}\right) d^{3}\vec{r}
+g\int n_{T}^{2}(\vec{r})d^{3}\vec{r}.
\label{eint}
\end{equation}

To obtain (\ref{eint}) we used the fact that 
$n_{T}(\vec{r})=\Sigma _{l\neq 0}N_{l}\phi _{l}^{2}(\vec{r})$ is the density distribution of the 
thermal atoms. Near the critical temperature,
the density distribution of the condensate is 
$n_{0}\left( \vec{r}\right) =N_{0}\phi _{0}^{2}\left( \vec{r}\right) =N_{0}\left( \frac{m\omega _{ho}}{\pi \hbar }\right) ^{3/2}exp\lbrack -\frac{m}{\hbar }\left( \omega _{x}x^{2}+\omega _{y}y^{2}+\omega _{z}z^{2}\right) \rbrack $.
The density distribution of the thermal atoms can be obtained using
Bose-Einstein distribution and the semiclassical approximation of the energy
level of a single particle. Near the critical 
temperature $\frac{k_{B}T}{\hbar\omega_{ho}} \propto N^{1/3}$.
In the available traps $N$
ranges from a few thousand to several millions, 
thus $\hbar \omega _{ho}<<k_{B}T_{c}^{0}$, {\it ie.} the energy level of the thermal atoms
can be approximated as continuous. In addition,
this means that the semiclassical approximation for the normal gas is expected to
work well on a wide range of temperatures \cite{RMP}.
In terms of Bose-Einstein distribution
$n_{T}\left( \vec{r}\right) =\int d^{3}\vec{p}\left( 2\pi \hbar \right) ^{-3}\left[ \exp \left( \beta \varepsilon \left( \vec{p},\vec{r}\right) \right) -1\right] ^{-1}$,
where $\varepsilon (\vec{p},\vec{r})=\frac{\vec{p}^{2}}{2m}+V_{ext}\left( 
\vec{r}\right) $ 
is the semiclassical energy in phase space and $\beta =1/k_{B}T$.
Density distribution of the thermal atoms in the coordinate space would be
$n_{T}(\vec{r})=\lambda _{T}^{-3}g_{3/2}\left( \exp \lbrack -V_{ext}(\vec{r}%
)/(k_{B}T)\rbrack \right)$, where
$\lambda _{T}=\left[ 2\pi \hbar ^{2}/\left( mk_{B}T\right) \right]
^{1/2}$ is the thermal wavelength. 
$g_{3/2}\left( z\right) $ belongs to the
class of functions $g_{\alpha }(z)=\Sigma _{n=1}^{\infty }z^{n}/n^{\alpha }$.
By integrating the semiclassical approximation $n_{T}(\vec r)$ over space
one obtain the condensate fraction 
$\xi =1-\left( \frac{T}{T_{c}^{0}}\right) ^{3}$, which does not account for finite-size effects.
However, in this paper the semiclassical approximation $n_{T}(\vec r)$ 
is used to calculate the interaction energy of the system, {\it ie.} used to calculate
the high-order correction due to interatomic interaction.
Thus the adoption of the
semiclassical approximation will omit only higher-order modification,
which is of the order of the multiplication of the corrections due to 
the interatomic interaction and finite-size effects.
Because the finite-size effects have been separated off in (\ref{main1}), we can give
finite-size effects although semiclassical approximation is used to calculate the correction due
to interatomic interaction.

Using the geometrical average of the oscillator frequencies $\omega _{ho}$, 
$n_{T}\left( r\right) =\lambda _{T}^{-3}g_{3/2}\left( \exp \left[ -m\omega _{ho}^{2}r^{2}/2k_{B}T\right] \right) $.
Introducing the width of the normal gas 
$R_{T}=\sqrt{2k_{B}T/m\omega _{ho}^{2}}$ and 
defining the rescaled variable through
$r=R_{T}\overline{r}$, we can obtain 
$n_{T}\left( \overline{r}\right) =\lambda _{T}^{-3}g_{3/2}\left( \exp \left[ -\overline{r}^{2}\right] \right) $.
Near the critical temperature $k_{B}T>>\hbar \omega _{ho}$, the classical 
Boltzmann distribution $n_{cl}\left( \overline{r}\right) \propto \lambda _{T}^{-3}\exp \left[ -\overline{r}^{2}\right] $
gives a zero-order approximation for the density distribution of the thermal cloud.
Thus we take the trial function of the form

\begin{equation}
n_{T}^{G}\left( \overline{r}\right) =\eta _{1}\lambda _{T}^{-3}\exp \left[ -\eta _{2}\overline{r}^{2}\right] 
\label {normal}
\end{equation}

\noindent where $\eta _{1}$ and $\eta _{2}$ are two dimensionless
variational parameters. Another reason for the adoption of the Gaussian trial function lies in the fact that
there is a factor $\exp \left[ -\overline{r}^{2}\right] $ in $n_{T}\left( \overline{r}\right) $.
$\eta _{1}$
and $\eta _{2}$ are\ determined by the requirement that $\int
|n_{T}^{G}\left( \vec{r}\right) -n_{T}\left( \vec{r}\right) |d^{3}\vec{r}$
obtain the minimum. Results of numerical calculations show that $\eta
_{1}=2.587$, $\eta _{2}=1.146$ and $\int |n_{T}^{G}\left( \vec{r}\right)
-n_{T}\left( \vec{r}\right) |d^{3}\vec{r}/\int n_{T}\left( \vec{r}\right)
d^{3}\vec{r}=0.0017$. So the Gaussian distribution $n_{T}^{G}(\vec{r})$
agrees very well with $n_{T}(\vec{r})$. We should note that the Gaussian distribution
$n_{T}^{G}\left( \vec{r}\right) $ is different from the classical Boltzmann distribution
because of the variational parameters $\eta _{1}$ and $\eta _{2}$.

Combining (\ref{eint}) and (\ref{normal}) we have

\begin{equation}
E_{int}=\frac{g}{2\left( \sqrt{2\pi }a_{ho}\right) ^{3}}N_{0}^{2}+
\left( 2-\frac{1}{2^{3/2}}\right) \eta _{1}g\lambda _{T}^{-3}N_{0}+\frac{\eta _{1}g\lambda _{T}^{-3}}{2^{3/2}}N,
\label{energy}
\end{equation}

\noindent where we have used the fact that $a_{ho}>>\lambda_{T}$ and 
$k_{B}T>>\hbar \overline{\omega }$ near the critical temperature. 

%%%%%%%%%%%%%%%%%%%%%%%%%%%%%%%
\subsection{shift of the critical temperature and condensate fraction near $T_{c}$}
%%%%%%%%%%%%%%%%%%%%%%%%%%%%%%%%%

We introduce a scaling parameter which accounts
for the role of the two-body repulsive interaction. This parameter $\theta$ 
is fixed by the ratio
between $gn_{T}\left( \vec{r}=0,T_{c}^{0}\right) $ and the critical temperature
$k_{B}T_{c}^{0}$. 
$\theta =\frac{gn_{T}\left( \vec{r}=0,T_{c}^{0}\right) }{k_{B}T_{c}^{0}}=\frac{\eta _{1}\sqrt{2/\pi }}{\zeta \left( 3\right) ^{1/6}}\frac{a}{a_{ho}}N^{1/6}$.
The scaling parameter $\theta$ can also be written in the form $\theta =0.65\eta ^{5/2}$, where 
$\eta =\mu \left( T=0\right) /k_{B}T_{c}^{0}
=\frac{\zeta \left( 3\right) ^{1/2}15^{2/5}}{2}\left( N^{1/6}\frac{a}{a_{ho}}\right) ^{2/5}$
is also an important scaling parameter accounting for the role of interatomic interaction
\cite{RMP,GOS}.
Obviously $\theta$ stands for high-order correction due to 
interatomic interaction, compared to the parameter $\eta$. From (\ref{main1}), (\ref{main2}),
and (\ref{energy}) we can obtain the equation on the condensate fraction.

\begin{equation}
1-\xi =\frac{t^{3}}{\zeta \left( 3\right) }g_{3}\left( z_{0}\right) +\frac{3\zeta \left( 2\right) }{2\zeta \left( 3\right) ^{2/3}}\frac{\overline{\omega }}{\omega _{ho}}t^{2}N^{-1/3}\label{near1},
\end{equation}

\begin{equation}
-\frac{\zeta \left( 3\right) ^{1/2}\theta N^{1/2}}{\eta _{1}t}\xi -\left( 2-\frac{1}{2^{3/2}}\right) \theta t^{1/2}+\ln z_{0}=0,
\label {near2}
\end{equation}

\noindent where we have used the reduced temperature $t=T/T_{c}^{0}$.
From (\ref{near1}) and (\ref{near2}) the condensate fraction is given by

\begin{equation}
\xi =\frac{1-t^{3}-4.51\frac{a}{a_{ho}}N^{1/6}t^{7/2}-\frac{3\zeta \left( 2\right) }{2\zeta \left( 3\right) ^{2/3}}\frac{\overline{\omega }}{\omega _{ho}}t^{2}N^{-1/3}}{1+1.16\frac{a}{a_{ho}}N^{2/3}t^{2}}.
\label{conden}
\end{equation}

The third term in the numerator of (\ref{conden}) represents the correction due to 
the interaction between atoms, while the last term in the numerator accounts for the
correction due to the finite-size effects. By setting $\xi =0$ in (\ref{conden}) one can obtain
the shift of the critical temperature,

\begin{equation}
\frac{\delta T_{c}^{0}}{T_{c}^{0}}=\frac{\delta T_{int}}{T_{c}^{0}}+\frac{\delta T_{finite}}{T_{c}^{0}}=-1.50\frac{a}{a_{ho}}N^{1/6}-\frac{\zeta \left( 2\right) }{2\zeta \left( 3\right) ^{2/3}}\frac{\overline{\omega }}{\omega _{ho}}N^{-1/3}.
\label{shift}
\end{equation}

The first term in (\ref{shift}) is the shift due to interatomic interaction. 
It agrees with the results based on the local density approximation \cite{RMP,GIO}.
Because of the denominator  in  (\ref{conden}), 
close to $T_{c}^{0}$ the condensate fraction will
increase slowly with decreasing the temperature. Especially, it shows that there is only
high-order correction of the scaling parameter $\eta$ due to interatomic interaction.
We should note that (\ref{conden}) holds only when $\frac{N_{0}a}{a_{ho}}<<1$.
The second contribution in (\ref{shift})
gives exactly the usual results due to the finite-size effects. 

%%%%%%%%%%%%%%%%%%%%%%%%%%%%%%%%
\section{condensate fraction of the system below $T_{c}$}
%%%%%%%%%%%%%%%%%%%%%%%%%%%%%%%%%

%%%%%%%%%%%%%%%%%%%%%%%%%%%%%%%%%
\subsection{lowest-order correction due to interatomic interaction and finite-size effects}
%%%%%%%%%%%%%%%%%%%%%%%%%%%%%%%%%%%

Below the critical temperature $\frac{N_{0}a}{a_{ho}}>>1$, $\it ie.$
$E_{int}>>E_{kin}$. In this case we can use the well-known Thomas-Fermi approximation.
With the Thomas-Fermi approximation the density profile of the condensate is
$n_{0}\left( \vec{r}\right) =\frac{\mu -V_{ext}\left( \vec{r}\right) }{g}\theta \left( \mu -V_{ext}\left( \vec{r}\right) \right) $.
The normalization condition on $n_{0}(\vec r)$ provides the relation between chemical 
potential and number of particles in the condensate: 
$\mu =\frac{\hbar \omega _{ho}}{2}\left( \frac{15N_{0}a}{a_{ho}}\right) ^{2/5}$.
From Gross-Pitaevskii (GP) equation \cite{GPE} and the well-known Virial theorem
the energy of the condensate turns out to be $E_{0}=\left( 5/7\right) \mu N_{0}$ \cite{RMP}.
Omitting atom-atom interactions
in the normal gas and the interaction between the condensate and normal gas, from 
(\ref{main1}) and (\ref{main2})
 
\begin{equation}
N_{0}=N-N_{T}=N-\Sigma_{n_{x},n_{y},n_{z}}^{\prime}\frac{1}{\exp \left[ \beta \left( \varepsilon _{n_{x}n_{y}n_{z}}-\mu \right) \right] -1}.
\label{low1}
\end{equation}

From (\ref{low1}) it is easy to obtain the following result

\begin{equation}
\xi =1-t^{3}-\frac{\zeta \left( 2\right) }{\zeta \left( 3\right) }\frac{\mu }{k_{B}T}-\frac{3\overline{\omega }\zeta \left( 2\right) }{2\omega _{ho}\zeta \left( 3\right) ^{2/3}}t^{2}N^{-1/3}\label{low2},
\end{equation}

\noindent where the last term in (\ref{low2}) is the correction due to finite-size effects \cite{FIN}.
Using the relation $\mu /k_{B}T=\eta \xi ^{2/5}/t$,
the third term in (\ref{low2}) gives exactly the usual lowest-order modification due to 
atom-atom interactions \cite{RMP,GIO,NAR}.

\begin{equation}
\xi =1-t^{3}-\frac{\zeta \left( 2\right) }{\zeta \left( 3\right) }\eta \xi ^{2/5}t^{2}-\frac{3\overline{\omega }\zeta \left( 2\right) }{2\omega _{ho}\zeta \left( 3\right) ^{2/3}}t^{2}N^{-1/3}.
\label{lowc}
\end{equation}

This proves the equivalence between
canonical ensemble and grand-canonical ensemble, even in the presence of interatomic
interactions. We will give high-order modification
in the following.
 
%%%%%%%%%%%%%%%%%%%%%%%%%%%%%%%%
\subsection{high-order modification due to interatomic interactions}
%%%%%%%%%%%%%%%%%%%%%%%%%%%%%%%%

Equation (\ref{lowc}) is obtained by omitting the interaction between thermal  atoms and 
interaction between thermal atoms and the condensate. When these two sorts of 
interactions are considered

\begin{equation}
E_{0}+E_{int}=\frac{5}{7}\mu N_{0}+2g\int n_{0}\left( \vec{r}\right) n_{T}\left( \vec{r}\right) d^{3}\vec{r}+g\int n_{T}^{2}\left( \vec{r}\right) d^{3}\vec{r}.
\label{high}
\end{equation}

The calculations of (\ref{high}) are straightforward. The results is given by

\begin{equation}
E_{0}+E_{int}=\frac{5\mu N_{0}}{7}+\frac{16\pi \eta _{1}\mu }{15}\frac{R_{\mu }^{3}}{\lambda _{T}^{3}}-\frac{16\pi \eta _{1}\eta _{2}}{35}\frac{R_{\mu }^{3}}{\lambda _{T}^{3}}\frac{\mu ^{2}}{k_{B}T}+\frac{\eta _{1}g\lambda _{T}^{-3}}{2^{3/2}}\left( N-N_{0}\right), 
\label{high1}
\end{equation}

\noindent where $R_{\mu }=\sqrt{2\mu /m\omega _{ho}^{2}}$ is the radius of the condensate. 
$R_{\mu }^{3}/\lambda _{T}^{3}=\alpha _{R\lambda }N^{1/2}t^{3/2}N_{0}^{3/5}$,
where $\alpha _{R\lambda }=0.294\left( a/a_{ho}\right) ^{3/5}$. From (\ref{high1})

\begin{equation}
-\beta \frac{\partial }{\partial N_{0}}\left( E_{0}+E_{int}\right) =-\left( 1-0.89\eta ^{5/2}t^{1/2}\right) \frac{\mu }{k_{B}T}-1.01\eta ^{5/2}t^{1/2}.
\label{high2}
\end{equation}

From (\ref{main1}), (\ref{main2}), and (\ref{high2}) we can obtain the condensate fraction
below the critical temperature,

\begin{equation}
\xi =1-t^{3}-\frac{\zeta \left( 2\right) }{\zeta \left( 3\right) }t^{3}\left[ \left( 1-0.89\eta ^{5/2}t^{1/2}\right)\frac{\eta \xi ^{2/5}}{t}+1.01\eta ^{5/2}t^{1/2}\right] -\frac{3\overline{\omega }\zeta \left( 2\right) }{2\omega _{ho}\zeta \left( 3\right) ^{2/3}}t^{2}N^{-1/3}.
\label{highc}
\end{equation}

In (\ref{highc}) the terms comprises $\eta ^{5/2}$ represents high-order correction to
the condensate fraction due to interatomic interaction. Omitting the terms comprises
$\eta ^{5/2}$, we can obtain the lowest-order correction (\ref{lowc}).
Essentially, (\ref{lowc}) and (\ref{highc}) are transcendental
equations on the condensate fraction. We can easily give
the numerical solution of the condensate fraction. In Fig.1 and Fig.2 we used
the experimental parameter by Ensher {\it et al.} \cite{ENS},
where the cloud
consists of $4\times 10^{4}$ atoms at the transition and 
$a/a_{ho}=5.4\times 10^{-3}$. According to Fig.2 the high-order
correction (\ref{highc}) agrees well with the condensate fraction (\ref{conden})
in the region $\frac{N_{0}a}{a_{ho}}<<1$.
From (\ref{conden}) and (\ref{highc}), close to $T_{c}^{0}$
the condensate fraction increase quite slowly with decreasing the temperature. However,
the lowest-order correction (\ref{lowc}) is rather different from the dashed line near the
critical temperature. The difference between (\ref{conden}) and (\ref{lowc}) lies in the
fact that there is only high-order correction due to interatomic interaction near the 
critical temperature, while there is only lowest-order correction in (\ref{lowc}). 
Fig.2 demonstrates the validity of (\ref{highc})
over the whole range of temperature below $T_{c}$.

In the preceding calculations we neglects the 
role of the interaction between the condensate and the thermal atoms on the 
density distributions of the condensate and the thermal atoms.
In the frame of Hartree-Fork model \cite{RMP,NAR},
the densities of the condensate
and the thermal component are given as follows.

\begin{equation}
n_{0}\left( \vec{r}\right) =\frac{\mu -V_{ext}\left( \vec{r}\right) -2gn_{T}\left( \vec{r}\right) }{g}\theta \left( \mu -V_{ext}\left( \vec{r}\right) -2gn_{T}\left( \vec{r}\right) \right), 
\label{density1}
\end{equation}

\begin{equation}
n_{T}\left( \vec{r}\right) =\lambda _{T}^{-3}g_{3/2}\left[ e^{-\left( v_{ext}\left( \vec{r}\right) +2g\left( n_{0}\left( \vec{r}\right) +n_{T}\left( \vec{r}\right) \right) -\mu \right) /k_{B}T}\right]. 
\label{density2}
\end{equation}

Using (\ref{main1}), (\ref{main2}), (\ref{high}) and (\ref{density1}), (\ref{density2})
and the usual iterative procedure 
we can obtain the numerical conclusion of the condensate fraction. The numerical 
conclusion is illustrated in Fig.3.

\section{condensate fraction beyond mean-field theory}

In the preceding calculations of the condensate fraction, GP equation is used
to obtain the energy $E_{0}$ of the condensate. GP equation is expected
to be valid if the system is dilute, {\it ie.} $n|a|^{3}<<1$. Obviously, correction to 
GP equation would modify the condensate fraction of the system. The first correction
to the mean-field approximation have been investigated by Timmermans, Tommasini, and
Huang 
\cite{TIM} and by Braaten and Nieto
\cite{BRA}. For large $N_{0}$, using
the local density approximation, the density distribution of the condensate is given by

\begin{equation}
n_{0}\left( \vec{r}\right) =\frac{\mu -V_{ext}\left( \vec{r}\right) }{g}-\frac{4m^{3/2}}{3\pi ^{2}\hbar ^{2}}\left[ \mu -V_{ext}\left( \vec{r}\right) \right] ^{3/2},
\label{GP1}
\end{equation}

\noindent with $\mu$ given by

\begin{equation}
\mu =\frac{\hbar \omega _{ho}}{2}\left( \frac{15N_{0}a}{a_{ho}}\right) ^{2/5}\left( 1+\sqrt{\pi a^{3}n\left( 0\right) }\right). 
\label{GP2}
\end{equation}

The parameter $a^{3}n\left( 0\right) $ can be directly expressed in terms of the relevant
parameters of the system. 
$a^{3}n\left( 0\right) =\frac{15^{2/5}}{8\pi }\left( N_{0}^{1/6}\frac{a}{a_{ho}}\right) ^{12/5}=0.0078\eta ^{6}\xi ^{2/5}$.
Using the relation $E_{0}=5N_{0}\mu /7$ and (\ref{main1}), (\ref{main2}) we can 
obtain the correction to the condensate fraction below $T_{c}$

$${\xi =1-t^{3}}$$

\begin{equation}
-\frac{\zeta \left( 2\right) }{\zeta \left( 3\right) }t^{3}\left[ \left( 1-0.89\eta ^{5/2}t^{1/2}+0.16\eta ^{3}\xi ^{1/5}\right) \frac{\eta \xi ^{2/5}}{t}+1.01\eta ^{5/2}t^{1/2}\right] -\frac{3\overline{\omega }\zeta \left( 2\right) }{2\omega _{ho}\zeta \left( 3\right) ^{2/3}}t^{2}N^{-1/3}.\label{GP3}
\end{equation}

The term $0.16\eta ^{3}\xi ^{1/5}$ is the correction due to the modification of the
GP equation.
It gives higher-order correction of the scaling parameter $\eta$, compared to the high-order
correction due to interatomic interactions.

Another correction to the condensate fraction beyond mean-field approximation is 
given by the quantum depletion of the condensate. There are atoms which do not occupy the
condensate at zero temperature because of quantum depletion effects. We can use local
density approximation to write the density of atoms out of the condensate. 
Timmermans, Tommasini, and Huang
\cite{TIM} gives 
$n_{out}\left( \vec{r}\right) =\left( 8/3\right) \left[ n\left( \vec{r}\right) a^{3}/\pi \right] ^{1/2}$.
Integration of $n_{out}\left( \vec{r}\right) $ yields the result

\begin{equation}
\xi _{out}=\int n_{out}\left( \vec{r}\right) d^{3}\vec{r}=\frac{5\sqrt{\pi }}{8}\sqrt{a^{3}n\left( 0\right)}.
\label{GP4}
\end{equation}

Combining (\ref{GP3}) and (\ref{GP4})

$$\xi =1-t^{3}-\frac{\zeta \left( 2\right) }{\zeta \left( 3\right) }t^{3}\left[ \left( 1-0.89\eta ^{5/2}t^{1/2}+0.16\eta ^{3}\xi ^{1/5}\right) \frac{\eta \xi ^{2/5}}{t}+1.01\eta ^{5/2}t^{1/2}\right] $$

\begin{equation}
-0.11\eta ^{3}\xi ^{1/5}-\frac{3\overline{\omega }\zeta \left( 2\right) }{2\omega _{ho}\zeta \left( 3\right) ^{2/3}}t^{2}N^{-1/3}.
\label{BEY}
\end{equation}

The term $-0.11\eta ^{3}\xi ^{1/5}$ is the correction due to quantum depletion. 

Compared to the finite-size effects, the
corrections to the mean-field approximation have a different dependence on the
parameter $N$ and $a/a_{ho}$. The corrections beyond mean field theory become 
larger than finite-size effects when $N$ is larger than about $10^{6}$. In Fig.4
we give numerical conclusion of (\ref{BEY}).
The corrections beyond mean field theory become important when the temperature
is much lower than the critical temperature.

\section{conclusion}

To conclude, the canonical ensemble is used to obtain the equations on the condensate 
fraction (\ref{main1}) and (\ref{main2}), which account for interaction effects.
Finite-size effects can be obtained simultaneously because these effects are
comprised in (\ref{main1}). (\ref{main1}) and (\ref{main2}) are used to investigate
the condensate fraction near and below the critical temperature. From the condensate
fraction near the critical temperature, we obtain the shift of the critical temperature
due to atom-atom interaction and finite-size effects simultaneously. In addition,
the analytical high-order correction due to interatomic interaction is obtained in this work.

\section*{Acknowledgments}
This work was supported by the Science Foundation of
Zhijiang College, Zhejiang University of Technology. This work was also
supported by the National Natural Science Foundation of China (19975019).

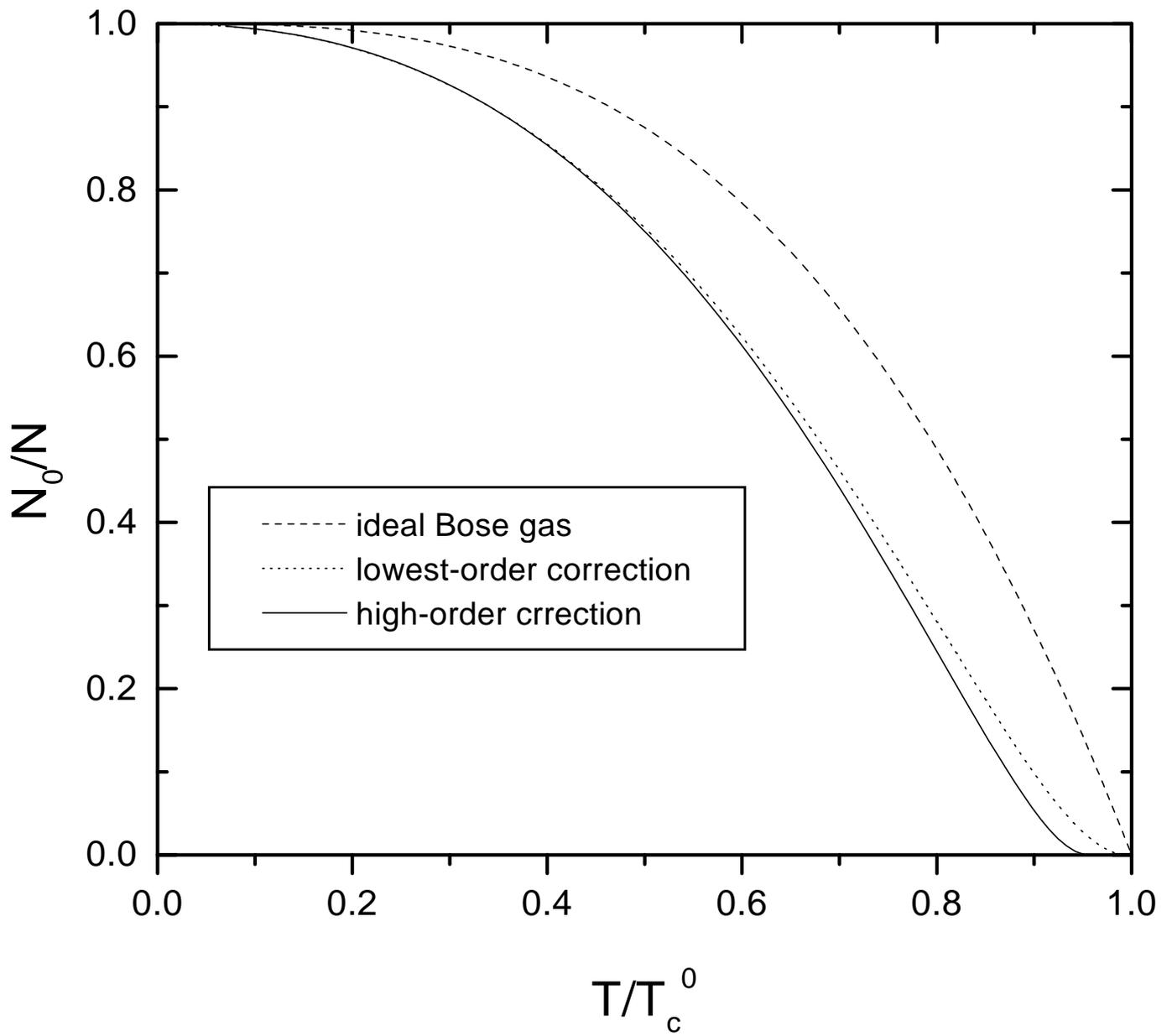
\begin{figure}
\caption{Theoretical prediction for the condensate fraction vs $T/T_{c}^{0}$ for
$N=4\times 10^{4}$ rubidium atoms in a trap with $a/a_{ho}=5.4\times 10^{-3}$.
The dashed line shows the noninteracting model in the thermodynamical limit.
The dotted line is obtained from the numerical solution of (\ref{lowc}),
which accounts for the lowest-order correction due to interatomic interaction.
The solid line is obtained from (\ref{highc}) which accounts for high-order correction.} 
\end{figure}

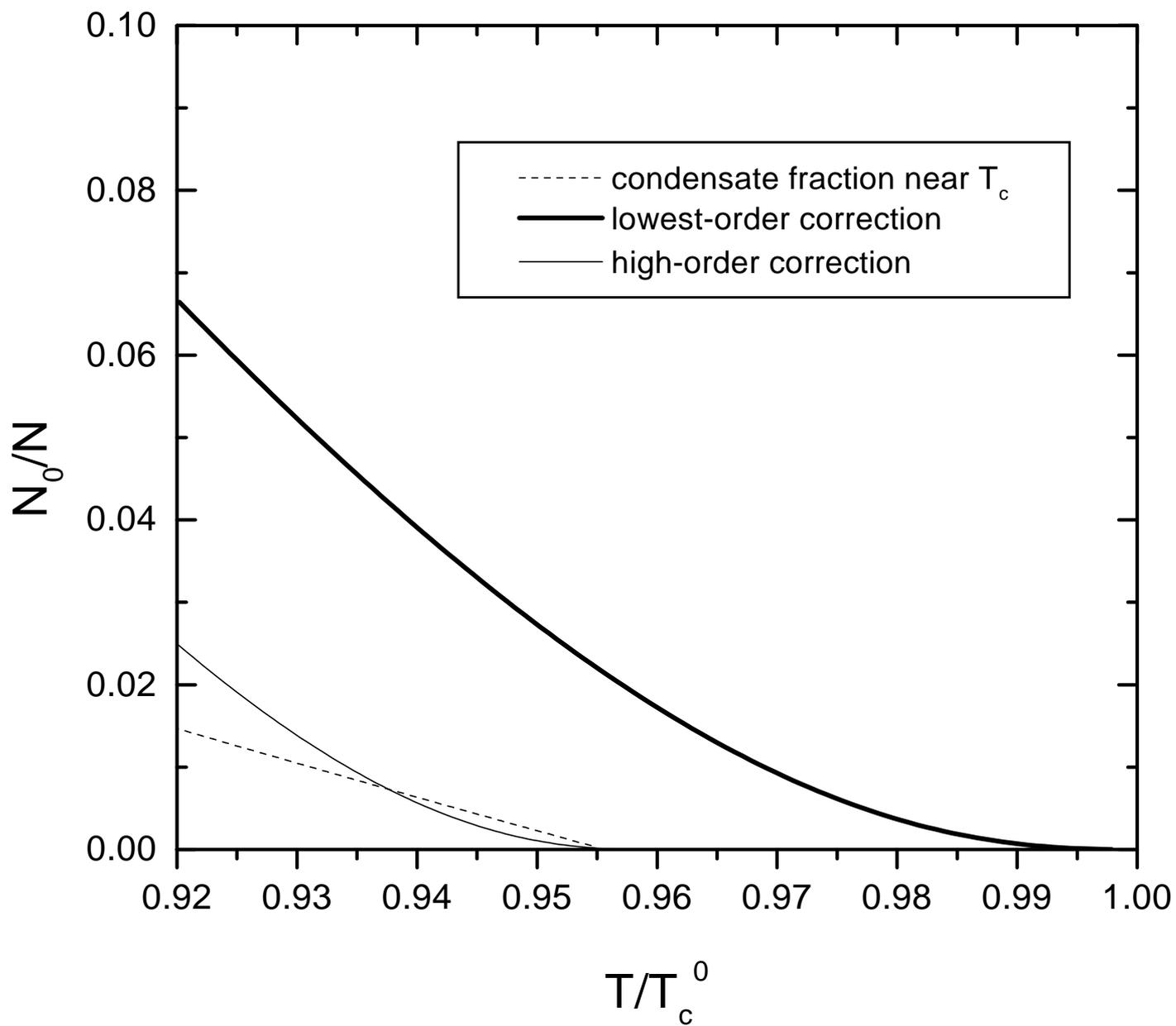
\begin{figure}
\caption{The upper solid line shows the lowest-order correction (\ref{lowc}) due to interatomic
interaction.
The lower solid line represents the high-order correction (\ref{highc}).
The dashed line shows the 
condensate fraction near $T_{c}$, which is obtained from (\ref{conden}).
Equation (\ref{conden}) holds when $\frac{N_{0}a}{a_{ho}}<<1$.
In the region $\frac{N_{0}a}{a_{ho}}<<1$, the lower solid line agrees well with the dashed
line. This demonstrates the validity of (\ref{highc}) over the whole range of temperature 
$T<T_{c}$, even close to $T_{c}^{0}$.}
\end{figure}

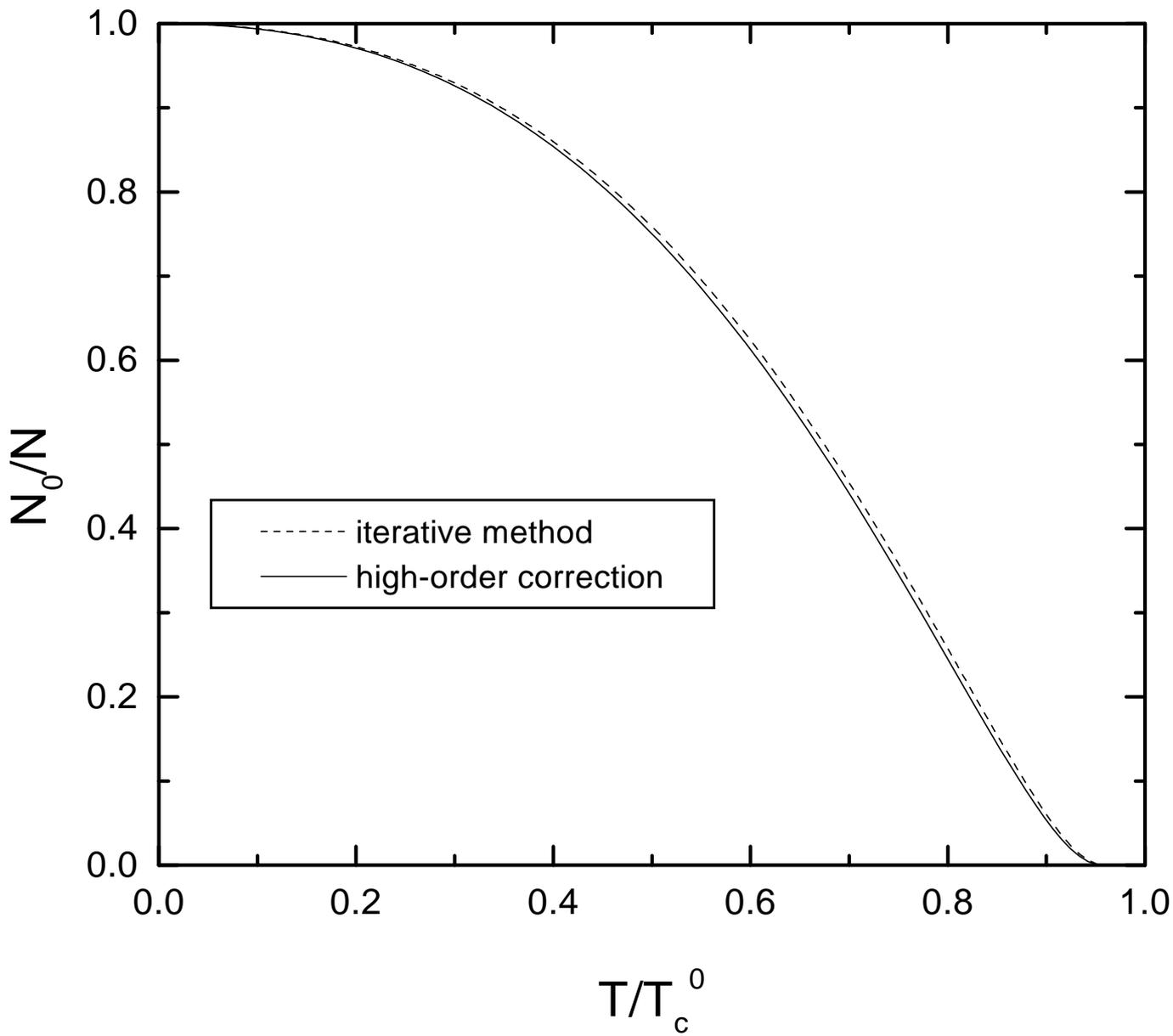
\begin{figure}
\caption{The solid line is obtained from (\ref{highc}). The dashed line is a self-consistent
numerical solution of (\ref{main1}), (\ref{main2}), (\ref{high}), and (\ref{density1}),
(\ref{density2}) using iterative method.}
\end{figure}

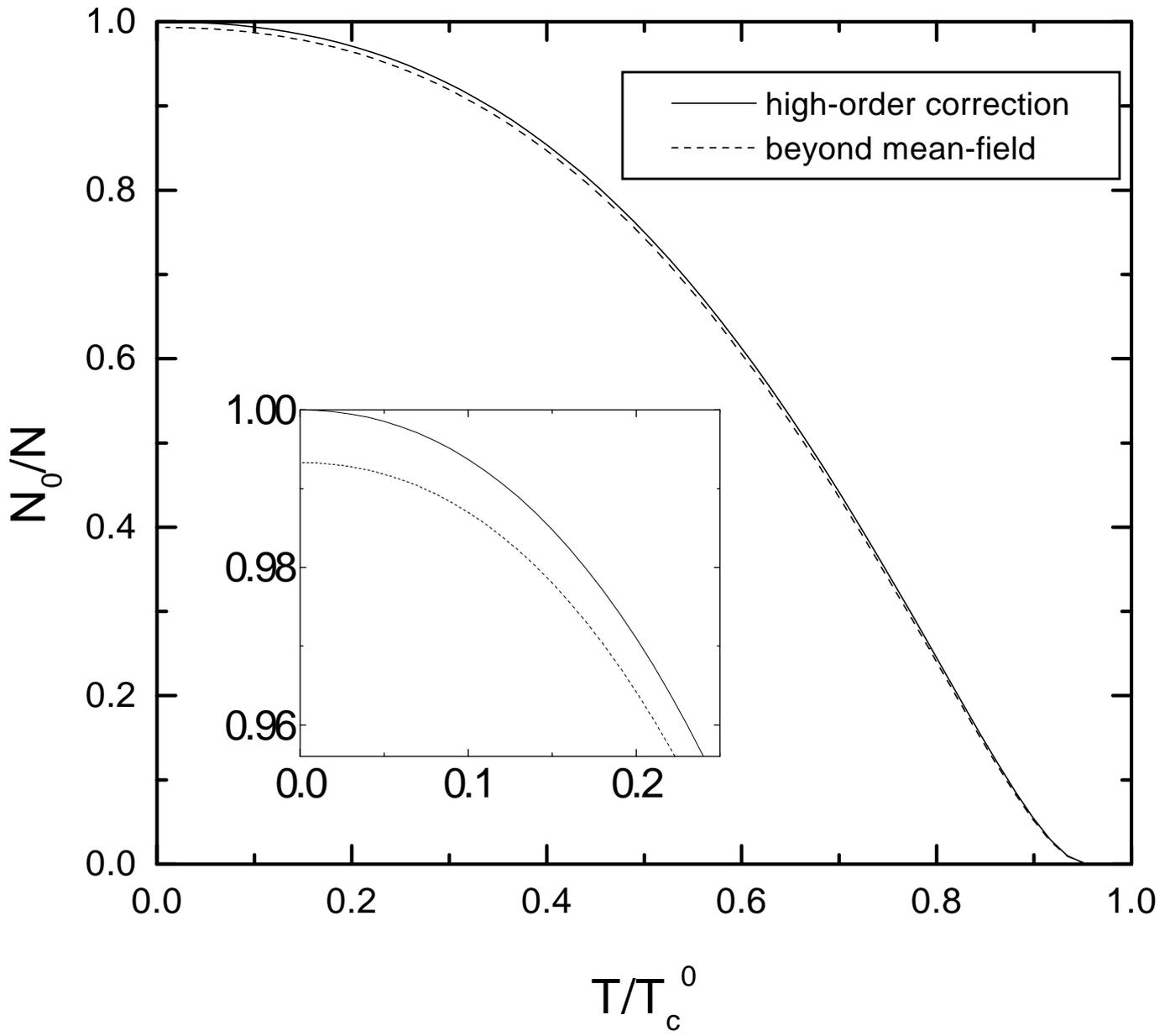
\begin{figure}
\caption{The solid line is obtained from (\ref{highc}) which accounts for high-order
correction due to interatomic interaction.
The dotted line is obtained from (\ref{BEY}), in which the corrections beyond mean field
theory are comprised. The corrections beyond mean-field theory become important
when the temperature is much lower than the critical temperature (shown in the inset).}
\end{figure}

\end{document}